\documentclass[12pt]{article}
\usepackage{amsmath}
\usepackage{amsfonts}
\usepackage{a4wide}
\usepackage{bbm}
\usepackage{heronb}
\newcommand{\I}{\text{i}}
\newcommand{\E}{\text{e}}

\newcommand{\re}[1]{(\ref{#1})}
\newcommand{\sta}[1]{{}^\star\! #1}
\newcommand{\Bcr}{B_{\text{cr}}}
\newcommand{\case}[2]{{\scriptstyle \frac{#1}{#2}}}

\begin{document}
\abovedisplayskip17pt plus2pt minus4pt
\abovedisplayshortskip14pt plus2pt minus4pt
\belowdisplayskip17pt plus2pt minus4pt
\belowdisplayshortskip14pt plus2pt minus4pt
\title{Applications of the Light Cone Condition for various perturbed
  Vacua.\thanks{Presented at the Fourth Workshop on ``Quantum Field
    Theory under the Influence of External Conditions'' at the
    University of Leipzig, Germany, September 14-18, 1998}}
\authors{Walter Dittrich and Holger Gies} 
  {Institut f\"ur theoretische Physik, Universit\"at T\"ubingen,\\ 
  Auf der Morgenstelle 14, 72076 T\"ubingen, Germany}

\maketitle

\begin{abstract}
We examine the propagation of light in the presence of various
modifications of the QED vacuum in the limit of low frequency. A
polarization summed and direction averaged-light cone condition is
derived from the equation of motion which arises from the effective
QED action. Several applications are given concerning vacuum
modifications caused by, e.g., strong fields, Casimir systems and
finite temperature.  \medskip

\noindent
Table of Contents \medskip

\noindent
1. Modification of the Vacuum Light Velocity $c$(=1) in Presence of
External Fields, Finite Temperature, Casimir Plates, etc. Examples.

\noindent
2. Light Cone Condition (LCC): Effective Action Approach.

\noindent
3. Applications of the LCC.
\end{abstract}

\section[]{Modification of the Vacuum Light Velocity $c$(=1) in
  Presence of External Fields, Finite Temperature, Casimir Plates,
  etc. Examples.} 

\label{sec1}
In this first chapter we want to list some examples which demonstrate
how the light velocity $c$(=1) becomes changed when a light ray
traverses a region which is disturbed by the presence of certain field
configurations, finite temperature environment, or Casimir plates.

\subsection*{(a) Weak electromagnetic fields.}

Let us focus on a pure constant applied magnetic field $\mathbf{B}$
which acts as a birefringent medium for the incoming photon beam. One
can distinguish between two polarization modes where either the
$\mathbf{e}$-field ($\bot$-mode) or the $\mathbf{h}$-field ($\|$-mode)
of the light wave points along the direction perpendicular to the
plane containing the $\mathbf{B}$-field and the wave propagation
direction , $(\mathbf{\hat{B},\hat{k}})$-plane. Then we obtain two
corresponding refractive indices $\left(
  \frac{k}{\omega} \right)_{\bot,\|}$ which are given
by \cite{1}:
\begin{eqnarray}
n_\bot &=&1+\frac{8\alpha^2}{45m^4}\left( 1+\frac{40}{9}
  \frac{\alpha}{\pi} \right) {B}^2\, \sin^2 \theta,\qquad
  \theta=<\!\!\!) (\mathbf{B,k}) \nonumber\\
n_\| &=&1+\frac{14\alpha^2}{45m^4}\left( 1+\frac{1315}{252}
  \frac{\alpha}{\pi} \right) {B}^2\, \sin^2 \theta. \label{1.1}
\end{eqnarray}
The change in the light velocity is correspondingly
$\left(v_{\|,\bot}=\frac{c(=1)}{n_{\|,\bot}} \right)$
\begin{eqnarray}
v_\bot &=&1-\frac{8\alpha^2}{45m^4}\left( 1+\frac{40}{9}
  \frac{\alpha}{\pi} \right) {B}^2\, \sin^2 \theta,\qquad
  \theta=<\!\!\!) (\mathbf{B,k}) \nonumber\\
v_\| &=&1-\frac{14\alpha^2}{45m^4}\left( 1+\frac{1315}{252}
  \frac{\alpha}{\pi} \right) {B}^2\, \sin^2 \theta. \label{1.2}
\end{eqnarray}
So, when we consider light of wavelength $\lambda$ travelling a path of
length  $L$ normal to the $\mathbf{B}$-field, the angular rotation of
the plane of polarization is given by \cite{2}
\begin{equation}
\Psi_{\text{QED}}=\frac{1}{15} \alpha \left( \frac{eB}{m^2} \right)^2
\frac{L}{\lambda} \left( 1+\frac{25}{4} \frac{\alpha}{\pi}
\right). \label{1.3}
\end{equation}
Using the two phase velocities \re{1.2} (to one-loop order only) the
average over polarization and direction is given by
\begin{eqnarray}
v&=& \frac{1}{4\pi} \int d\Omega\, \frac{1}{2} \bigl(v_\bot +b_\|
\bigr) =1-\frac{22}{135} \frac{\alpha^2}{m^4} B^2 \label{1.4}\\
\text{or}\quad \delta v&=&-\frac{44}{135} \frac{\alpha^2}{m^4} \,
u,\qquad u=\frac{1}{2} B^2. \label{1.5}
\end{eqnarray}

\subsection*{(b) Temperature-induced velocity shift.}

Here the velocity of soft photons moving in a photon gas at low
temperature $T\ll m$ is given by \cite{3}
\begin{equation}
v=1-\frac{44\pi^2}{2025} \alpha^2 \frac{T^4}{m^4}. \label{1.6}
\end{equation}

\subsection*{(c) Casimir vacuum.}

For a photon propagating perpendicular to Casimir plates with distance
$a$ one obtains \cite{4}
\begin{equation}
v=1+\frac{11}{(90)^2} \frac{\alpha^2}{m^4} \frac{\pi^2}{a^4} >1\qquad
! \label{1.7}
\end{equation}\bigskip

%\subsection*{(d) Gravitational background \cite{5}.}
%\begin{equation}
%v=1+\frac{11}{45} \frac{\alpha G_{\text{N}}}{m^2} \bigl(\rho+p\bigr)
%>1\qquad ! \label{1.8}
%\end{equation}
%    INSTEAD:
Further velocity shifts can be found for photons moving in a
gravitational background \cite{5}. Observe that all the examples are
low-energy phenomena.

Covering all these aforementioned cases Latorre, Pascual and Tarrach
\cite{6} presented an intriguing general, so-called ``unified''
formula. They claimed that the polarization and direction-averaged
velocity shift is related to the (renormalized) background energy
density $u$ with a ``universal'' numerical coefficient
\begin{equation}
\delta v=-\frac{44}{135} \frac{\alpha^2}{m^4}\, u. \label{1.9}
\end{equation}

That all these cases can only be of limited validity becomes clear
when looking at the strong $\mathbf{B}$-field regime
$(B>\Bcr=\frac{m^2}{e})$ where one finds \cite{7}
\begin{equation}
v^2=1-\frac{\alpha}{4\pi} \sin^2 \theta \left[ \frac{2}{3}
  \frac{B}{\Bcr} +{\cal O}(1) +{\cal O}\left( \frac{\Bcr}{B} \ln
  \frac{B}{\Bcr} \right) \right]. \label{1.10}
\end{equation}
Also of limited value is Shore's conjecture \cite{8} that the
``universal'' coefficient in \re{1.9} can be related to the trace
anomaly of the energy momentum tensor:
\begin{equation}
\langle T^{\alpha}{}_{\alpha}\rangle_{\text{E.M.}}=-4
  \Biggl[\underbrace{\frac{8}{45}\frac{\alpha^2}{m^4}}_{\sim 
  \delta v_\bot}\,\left(\frac{1}{4}F_{\mu\nu}F^{\mu\nu}\right)^2 
  +\underbrace{\frac{14}{45}\frac{\alpha^2}{m^4}}_{\sim \delta v_\|}\,
  \left(\frac{1}{4}\sta{F}_{\mu\nu}F^{\mu\nu}\right)^2\Biggr]. \label{1.11}
\end{equation}
It is our goal to rederive some of the former results and generalize
the ``unified formula'' of Latorre, Pascual and Tarrach. We confine
ourselves to the case of non-trivial vacua modified by QED phenomena.

\section{Light Cone Condition (LCC): Effective Action Approach
  \protect\cite{9}.} 
\label{sec2}
\setcounter{equation}{0}

Here are the essential assumptions that are needed for a description
of light propagation in various perturbed vacua employing the
effective action approach:

\noindent
(1) The propagating photons with field strength $f^{\mu\nu}$ are
considered to be soft: $\frac{\omega}{m}\ll 1$. 

\noindent
(2) The vacuum modification is homogeneous in space and time.

\noindent
(3) Vacuum modifications caused by the propagating light itself are
negligible. 

Let us now turn to pure electromagnetic vacuum modifications where the
dynamical building blocks of the effective action which respect
Lorentz and gauge invariance are given by
\begin{eqnarray}
F^{\mu\nu}&=&\partial^\mu A^\nu -\partial^\nu A^\mu, \label{2.1}\\
\sta{F}^{\mu\nu}&=&\frac{1}{2}\epsilon^{\mu\nu\alpha\beta}
F_{\alpha\beta}. \label{2.2}
\end{eqnarray}
The lowest-order linearly independent scalars (pseudo-scalars) are
\begin{eqnarray}
x&:=& \frac{1}{4}F_{\mu\nu}F^{\mu\nu} =\frac{1}{2} \bigl(
\mathbf{B^2-E^2} \bigr)\qquad\,\bigl(={\cal F}\bigr), \label{2.2a}\\
y&:=& \frac{1}{4}\sta{F}_{\mu\nu}F^{\mu\nu} =-\mathbf{E\cdot B}
\qquad\qquad \bigl(={\cal G}\bigr). \label{2.3} 
\end{eqnarray}
Hence the Maxwell Lagrangian can be written as ${\cal
  L}_{\text{M}}=-x$. The fundamental algebraic relations \cite{10},
\begin{eqnarray}
F^{\mu\alpha} F^\nu_{\,\,\,\alpha} -\, ^\star\! F^{\mu\alpha}\,
  ^\star\! F^{\nu}_{\,\,\,\alpha} &=& 2\, x\, g^{\mu\nu}\,
  ,\nonumber\\ 
F^{\mu\alpha}\, ^\star\! F^{\nu}_{\,\,\,\alpha} =\, ^\star\!
  F^{\mu\alpha} F^\nu_{\,\,\,\alpha} &=& y\, g^{\mu\nu},\label{2.4}
\end{eqnarray}
help to verify (i) the vanishing of odd-order invariants and (ii) that
invariants of arbitrary order can be reduced to expressions only
involving $x^ny^m$; $n,m=0,1,2\dots$ . Besides, note that parity
invariance demands for $m$ to be even. The corresponding Lagrangian
becomes then simply a function of $x$ and $y$:
\begin{equation}
{\cal L}={\cal L}(x,y). \label{2.5}
\end{equation}
From here we obtain the equations of motion by variation,
\begin{equation}
0=\partial_\mu \frac{\partial {\cal L}}{\partial (\partial_\mu
  A_\nu)} -\frac{\partial {\cal L}}{\partial A_\mu} 
=\partial_\mu \bigl( \partial_x {\cal L}\, F^{\mu\nu} +\partial_y
  {\cal L}\, ^\star\! F^{\mu\nu} \bigr), \label{2.6}
\end{equation}
or, after taking advantage of the Bianchi identity while moving
$\partial_\mu$ to the right,
\begin{equation}
0=(\partial_x {\cal L})\, \partial_\mu F^{\mu\nu} +\left(\frac{1}{2}
M^{\mu\nu}_{\alpha\beta}\right)\, \partial_\mu F^{\alpha\beta},
\label{2.7}
\end{equation}
where
\begin{equation}
M^{\mu\nu}_{\alpha\beta}:= F^{\mu\nu}F_{\alpha\beta}\, (\partial^2_x
  {\cal L}) +\, ^\star\! F^{\mu\nu}\, \sta{F}_{\alpha\beta}\,
  (\partial^2_y {\cal L}) +\partial_{xy}{\cal L}\, \bigl( F^{\mu\nu}\,
  \sta{F}_{\alpha\beta}+\, \sta{F}^{\mu\nu} F_{\alpha\beta} \bigr)
  .\label{2.8} 
\end{equation}
What follows is a 5-step calculation consisting of\medskip

\noindent
(1) \qquad $F^{\mu\nu} \longrightarrow
\underbrace{F^{\mu\nu}}_{\text{const. background field}}
+\underbrace{f^{\mu\nu}}_{\text{propag. light wave}}$ : $\quad
\partial_\mu F^{\kappa\lambda} \longrightarrow \partial_\mu
f^{\kappa\lambda}$\medskip

\noindent
(2) \qquad $f^{\mu\nu}=k^\mu a^\nu -k^\nu a^\mu =a(k^\mu \epsilon^\nu
-k^\nu \epsilon^\mu)$ where $k_\mu\epsilon^\mu=0$ in the Lorentz
gauge\medskip 

\noindent
(3) \qquad Average over polarization states: $\sum_{\text{pol.}}
\epsilon^\beta \epsilon^\nu \to g^{\beta\nu}$.\medskip

These intermediate steps yield instead of \re{2.7}
\begin{equation}
0=2(\partial_x {\cal L})\, k^2 +M^{\mu\nu}_{\alpha\nu}\, k_\mu k^\alpha 
.\label{2.9}
\end{equation}
Equation \re{2.9} already represents a light cone condition and
indicates that the familiar $k^2=0$ relation will in general not hold
for arbitrary Lagrangians.\medskip

\noindent
(4) \qquad Using the fundamental relations \re{2.4} together with the
Maxwell energy-momen\-tum tensor, 
\begin{equation}
T^\mu{}_\alpha=F^{\mu\nu}F_{\alpha\nu}-x\, \delta^\mu_\alpha,
\label{2.10}
\end{equation}
we obtain
\begin{equation}
M^{\mu\nu}_{\alpha\nu}=2\left[\frac{1}{2}T^\mu{}_ \alpha
  (\partial^2_x +\partial^2_y){\cal L} +\delta^\mu_\alpha \left(
  \frac{1}{2} x (\partial^2_x -\partial^2_y){\cal L} +y
  \partial_{xy}{\cal L}\right)\right] .\label{2.11}
\end{equation}\medskip

\noindent
(5) \qquad Finally we need the VEV of the energy-momentum tensor 
\begin{equation}
\langle T^{\mu\nu}\rangle_{xy}=-T^{\mu\nu} (\partial_x {\cal L})
+ g^{\mu\nu}\, ({\cal L} -x\partial_x {\cal L} -y\partial_y {\cal
  L})=:\frac{2}{\sqrt{-g}} \frac{\delta \Gamma}{\delta g_{\mu\nu}},
\label{2.12} 
\end{equation}
where $\Gamma:=\int d^4x\, \sqrt{-g}\, {\cal L}$ denotes the effective
action.\medskip

After these 5 steps we end up with the desired LCC:
\begin{equation}
k^2\, =\, Q\, \langle T^{\mu\nu}\rangle_{xy}\, k_\mu k_\nu
,\label{2.13} 
\end{equation}
with
\begin{equation}
Q=\frac{\frac{1}{2} (\partial^2_x +\partial^2_y){\cal L}}
{{ \Bigl[ \!(\partial_x\!{\cal L})^2\!+(\partial_x\!{\cal
    L})\!\bigl(\!\frac{x}{2} (\partial^2_x\! -\partial^2_y)+y
  \partial_{xy}\!\bigr)\!{\cal L}\!+\frac{1}{2}\! (\partial^2_x
  \!+\partial^2_y){\cal L}(1\! -x\partial_x \!-y\partial_y)\!
  {\cal  L} \Bigr]}}. \label{2.14}
\end{equation}
Now we want to extend the LCC to arbitrary vacuum disturbances, not
only of electromagnetic field type. Let us therefore parametrize the
additional vacuum modifications by the label $z$ and write instead of
\re{2.13}
\begin{eqnarray}
k^2\, &=&\, _z\langle 0|\,Q\, \langle T^{\mu\nu}\rangle_{xy}\,
|0\rangle_z \, k_\mu k_\nu \nonumber\\
&=&\sum_i  {}_z\!\langle 0|\,Q\,|i\rangle_z
  \,{}_z\!\langle i| \langle T^{\mu\nu}\rangle_{xy}\, |0\rangle_z \,
  k_\mu k_\nu. \label{2.15}
\end{eqnarray}
In the sequel, we assume the vacuum to behave as a passive medium
which leads to 
\begin{equation}
{}_z\langle 0|\,Q\,|i\rangle_{z}=\langle Q\rangle_z\, \delta_{0i}\,
  .\label{2.16} 
\end{equation}
This equation states that the vacuum exhibits no back-reaction caused
by the EM fields while switching on $z$. One can think of this
approximation as a kind of adiabatic or Born-Oppenheimer
approximation.

$Q$ depends functionally on ${\cal L}(x,y)\equiv {\cal
  L}^{\text{eff}}(A^{\text{ext}})$, which is, as usual, defined via
the functional integral over the fluctuating fields:
\begin{displaymath}
\E^{\I \int d^4q\, {\cal L}^{\text{eff}}(A^{\text{ext}})} =\int \bigl[
d\psi d\bar{\psi} dA\bigr]\, \exp\Bigl\{ \I S_{\text{QED}} \bigl[
\psi,\bar{\psi},A;A^{\text{ext}} \bigr] \Bigr\}
=Z\bigl[A^{\text{ext}}\bigr].
\end{displaymath}
From here we obtain for the definition of $\langle {\cal
  L}^{\text{eff}} \rangle_z\equiv {\cal L}(x,y;z)$
\begin{displaymath}
\E^{\I \int d^4q\, {\cal L}(x,y;z)} =\int\limits_z \bigl[
d\psi d\bar{\psi} dA\bigr]\, \exp\Bigl\{ \I S_{\text{QED}} \bigl[
\psi,\bar{\psi},A;A^{\text{ext}} \bigr]\Bigr|_z \Bigr\}
=Z\bigl[A^{\text{ext}}\bigr]\Bigr|_z.
\end{displaymath}
E.g., if the modification $z$ imposes boundary conditions on the
fields, as is the case for the Casimir effect or the thermalization of
photons or fermions in loop graphs, the functional integral has to be
taken over the fields which satisfy these boundary
conditions. Therefore, taking the VEV of $Q$ defines the new effective
Lagrangian characterizing the complete modified vacuum:
\begin{equation}
\langle Q\rangle_z=\langle Q({\cal L}(x,y))\rangle_z=Q({\cal
  L}(x,y;z))\, .\label{2.17}
\end{equation}
Here, then, is the LCC for arbitrary homogeneous modified vacua:
\begin{equation}
k^2=Q(x,y,z)\, \langle T^{\mu\nu} \rangle_{xyz}\, k_\mu k_\nu\,
.\label{2.18}
\end{equation}
If we choose a special Lorentz frame, 
\begin{equation}
\bar{k}^\mu=\frac{k^\mu}{|\mathbf{k}|} =\left(\frac{k^0}{|\mathbf{k}|}
  ,\mathbf{\hat{k}} \right) =:(v, \mathbf{\hat{k}}) ,\label{2.19}
\end{equation}
we obtain for Eq. \re{2.18}:
\begin{equation}
v^2=1-Q\, \langle T^{\mu\nu}\rangle \bar{k}_\mu\bar{k}_\nu
.\label{2.20}
\end{equation}
This LCC is a generalization of the ``unified formula'' of Latorre,
Pascual and Tarrach \cite{6}.

If we, furthermore, average over propagation directions, i.e.,
integrate over $\mathbf{\hat{k}}\in S^2$ and assume $Q\langle
T^{00}\rangle, \langle T^\alpha{}_\alpha\rangle\ll 1$, we obtain:
\begin{equation}
v^2=1-\frac{4}{3}\, Q\, \langle T^{00}\rangle= 1-\frac{4}{3}\, Q\, u
.\label{2.21}
\end{equation}
At this stage let us return to Shore's conjecture \cite{8} which
suggests a deeper connection between the velocity shift and the trace
anomaly. From \re{2.12}, we can read off the relation
\begin{eqnarray}
\langle T^{\alpha}{}_{\alpha}\rangle&=&4 ({\cal L} -x \partial_x
{\cal L} -y \partial_y {\cal L} )\label{2.22}\\
&\stackrel{\text{H.E.}}{=}&-4\left[\underbrace{\frac{8}{45}
    \frac{\alpha^2}{m^4}}_{\sim \delta v_\bot}\,x^2 
  +\underbrace{\frac{14}{45}\frac{\alpha^2}{m^4}}_{\sim \delta v_\|}\,
  y^2\right]. \label{2.23}
\end{eqnarray}
So there is a relation between the trace anomaly and the velocity
shift for different polarization states. Notice, however, that this
result is tied to using the Heisenberg-Euler Lagrangian in the weak
field limit. In general it is the $Q$-factor that is linked to
$\langle T^\alpha{}_\alpha\rangle$, as can be seen by writing the
numerator of \re{2.14} in the form
\begin{equation}
\text{num}(Q)=\frac{1}{2}(\partial^2_x+\partial^2_y){\cal
  L}=-\frac{1}{2}\left(\frac{y}{x}+\frac{x}{y}\right)
  \partial_{xy}{\cal L} -\frac{1}{8} \left(\frac{1}{x}\partial_x
  +\frac{1}{y}\partial_y \right) \langle
  T^{\alpha}{}_{\alpha}\rangle \label{2.24} 
\end{equation}
and only for the special structure of the H.E.-Lagrangian where
$\partial_{xy} {\cal L}=0$ is Shore's conjecture true. Referring to
the right-hand side of \re{2.20}, the value and sign of the velocity
shift result from the competition between the VEV of the
energy-momentum tensor and the $Q$-factor. For small corrections to
the Maxwell Lagrangian, 
\begin{equation}
{\cal L}={\cal L}_{\text{M}}+{\cal L}_{\text{c}}, \label{2.25}
\end{equation}
we have $\text{denom}(Q)=1+{\cal O}({\cal L}_{\text{c}})$, which
reduces Eq. \re{2.14} to
\begin{equation}
Q\simeq \frac{1}{2}(\partial^2_x+\partial^2_y){\cal L} \quad
\Longrightarrow \mathbf{\nabla}^2 {\cal L}=2\,Q \, .\label{2.26}
\end{equation}
Because of the similarity to the (2D) Poisson equation, from now on we
will call $Q$ the effective action charge in field space. The
classical vacuum ${\cal L}_{\text{M}}=-x$ is uncharged and hence
$v=1$. As we will soon demonstrate, the pure QED vacuum has a small
positive charge at the origin in field space $(x=0=y)$.

\section{Applications of the LCC.}
\label{sec3}
\setcounter{equation}{0}

\subsection*{A. Weak EM fields.}

Let us start with the two-loop Heisenberg-Euler Lagrangian \cite{11},
\begin{equation}
{\cal L}= -x+c_1\, x^2+ c_2\, y^2,\label{3.1}
\end{equation}
with
\begin{eqnarray}
c_1&=&\frac{8\alpha^2}{45 m^4} \left( 1+\frac{40}{9}\frac{\alpha}{\pi}
  \right),\nonumber\\
c_2&=&\frac{14\alpha^2}{45 m^4} \left(
  1+\frac{1315}{252}\frac{\alpha}{\pi} \right) . \label{3.2}
\end{eqnarray}
Then we obtain for the effective action charge
\begin{equation}
Q=\frac{1}{2} \bigl( \partial_x^2+\partial_y^2\bigr) {\cal L}=c_1+c_2
,\label{3.3}
\end{equation}
so that Eq. \re{2.21} immediately yields:
\begin{equation}
v=1-\frac{4\alpha^2}{135 m^4} \left(11+\frac{1955}{36}
  \frac{\alpha}{\pi} \right) \!\left[ \frac{1}{2} \bigl(
  \mathbf{E}^2+\mathbf{B}^2 \bigr) \right].\label{3.4}
\end{equation}

\subsection{B. Strong fields}
Here we begin with Schwinger's proper-time expression \cite{10} --
still valid for arbitrary constant field strength --
\begin{eqnarray}
{\cal L}\!=\!-x-\frac{1}{8\pi^2}\!\!
  \int\limits_0^{\I\infty}
  \!\!\frac{ds}{s^3}\E^{\!-m^2\!s}\!\biggl[(es)^2|y|
  \coth\!\Bigl(\!es\bigl(\!
  \sqrt{x^2\!+\!y^2}\!+x\bigr)\!^{\case{1}{2}}\!\Bigr) 
  &&\!\!\!\!\!\!\!\!\!\!\cot\!\Bigl(\! es\bigl(\!\sqrt{x^2\!+\!y^2}
  \!-x\bigr)\!^{\case{1}{2}}\!\Bigr)\!\nonumber\\
  &&\qquad\,-\frac{2}{3}(es)^2
  x-1\!\biggr]. \label{3.5}
\end{eqnarray}
The complete formula for the effective action charge for a purely
magnetic background field can be performed analytically and yields
$\left( h=\frac{\Bcr}{2B} \right)$:
\begin{eqnarray}
Q(h)=\frac{1}{2B^2}\frac{\alpha}{\pi}\biggl[&& \!\!\!\!\!\!\!\!\!\!
  \Bigl(\! 2h^2\! -\frac{1}{3}\Bigr) \Psi (1\!+\!h) -h -3h^2-4h\ln
  \Gamma (h) \nonumber\\
&&\qquad+2h\ln 2\pi +\frac{1}{3}+4\zeta'(-1,h) +\frac{1}{6h} \biggr] 
  .\label{3.6}
\end{eqnarray}
For strong fields, the last term in \re{3.6}, $\propto \frac{1}{6h}
\propto B$, dominates the expression in the square brackets. Hence,
the effective action charge decreases with
\begin{equation}
Q(B)\simeq \frac{1}{6}\frac{\alpha}{\pi} \frac{1}{\Bcr}
\frac{1}{B} ,\qquad\qquad \text{for}\quad B\to \infty. \label{3.7}
\end{equation}
Finally, the light cone condition \re{2.20} yields, for strong
magnetic background fields for which $B\stackrel{>}{\sim}\Bcr$.
\begin{eqnarray}
v^2=1-\frac{\alpha}{\pi} \frac{\sin^2 \theta}{2} &&\!\!\!\!\!\!\!\!
  \biggl[\!\Bigl(\case{B_{\text{cr}}^2}{2B^2}\!-\!\case{1}{3}\Bigr)
  \psi (1\!+ \!\case{B_{\text{cr}}}{2B})-\case{2B_{\text{cr}}}{B}\ln
  \Gamma (\case{B_{\text{cr}}}{2B}) -\case{3B_{\text{cr}}^2}{4B^2}  
  \label{3.8}\\
&&-\case{B_{\text{cr}}}{2B}+\case{B_{\text{cr}}}{B}\ln 2\pi
  +\!\frac{1}{3}\!+4 \zeta '(-1,\case{B_{\text{cr}}}{2B})
  +\case{B}{3B_{\text{cr}}}  \biggr], \nonumber\\
\longrightarrow 1-\frac{\alpha}{4\pi}  &&\!\!\!\!\!\!\!\!\sin^2 \theta
  \frac{2}{3} \frac{B}{\Bcr} \qquad\text{for} \quad
  B\stackrel{>}{\sim}\Bcr, \label{3.9} \\
\text{e.g.,}\quad\delta v\simeq 9.58..\cdot 10^{-5}&& \qquad\qquad
  \quad\,\,\text{at} \quad B=\Bcr =\frac{m^2}{e} , \label{3.10}
\end{eqnarray}
%Although the velocity shift increases proportional to the magnetic
%field for large $B$, the total amount of velocity shift remains
%comparably small,
%\begin{equation}
%\delta v\simeq 9.58\dots\cdot 10^{-5} \quad \text{at} \quad B=\Bcr
%=\frac{m^2}{e} , \label{3.10}
%\end{equation}
%for strong fields, consistent with the one-loop approximation, i.e.,
%$\frac{B}{\Bcr} < \frac{\pi}{\alpha}\simeq 430$. 
% INSTEAD:
At least on a formal stage, even the $B\to\infty$ limit can be
taken. For this, the complete structure of the effective action charge
$Q$ in Eq. \re{2.14} has to be taken into account as well as the
phase velocity dependence of the product $\langle T^{\mu\nu}\rangle
\bar{k}_\mu \bar{k}_\nu$ in Eq. \re{2.21}. The result for the velocity
square reads:
\begin{equation}
v^2=1-\sin^2 \theta +\frac{6\pi}{\alpha} \frac{\Bcr}{B} \sin^2 \theta+
\dots, \qquad\quad \text{for} \quad \frac{B}{\Bcr}\to
\infty. \label{3.10a} 
\end{equation}
In this limit, we find that a photon propagation perpendicular to the
magnetic field is strongly suppressed, while a propagation along the
magnetic field lines obtains no modifications. However, we must
remember that we calculated the polarization-summed average velocity,
and so the true modes of propagation might be different. Nevertheless,
this result shows similarities to the propagation of photons along the
magnetic field lines in a plasma (Alfv\'{e}n-mode propagation).
\subsection*{C. Casimir Vacua (Scharnhorst effect \cite{4}).}

In accordance with experimental facilities, the plate separation $a$
is treated as a macroscopic parameter $(a\propto \mu\text{m})$. In
order not to violate the soft photon approximation
$m\gg\frac{1}{\lambda}$, the photon wavelength has to obey $\lambda\ll
a$. Only then can we treat the Casimir region as a (macroscopic)
medium. 

Accepting these assumptions, $Q$ is simply given by
\begin{eqnarray}
Q&=&Q(x,y,a)\Big|_{x=0=y}=Q(x,y,0)\Big|_{x=0=y}+\underbrace{\Delta
  Q(x,y,a)}_{\sim \E^{-ma}} \Big|_{x=0=y}\stackrel{ma\gg 1}{\simeq}
  Q(x,y)\Big|_{x=0=y} \nonumber\\
&=&c_1+c_2=\frac{2}{45} \frac{\alpha^2}{m^4} \left( 11+\frac{1955}{36}
  \frac{\alpha}{\pi} \right). \label{3.11}
\end{eqnarray}
Together with $\langle T^{\mu\nu}\rangle$, as given in Ref. \cite{12},
\begin{equation}
\langle T^{\mu\nu}\rangle =\frac{\pi^2}{720a^4} 
\left( \begin{array}{cccc}
       -1&  &  &  \\
         & 1&  &  \\
         &  & 1&  \\
         &  &  &-3 \end{array} \right) \label{3.12}
\end{equation}
and Eq. \re{2.20}, we obtain for the propagation of light
perpendicular to the plates the superluminal phase and group velocity:
\begin{equation}
v=1+\frac{1}{(90)^2} \frac{\alpha^2}{m^4}\left(11+\frac{1955}{36}
  \frac{\alpha}{\pi} \right)\frac{\pi^2}{ a^4}. \label{3.13}
\end{equation}
Equation \re{3.13} represents the two-loop corrected version
of Scharnhorst's formula.\footnote{Potentially, there might be a
  further contribution to the two-loop correction, since the Casimir
  boundary conditions for the radiative photon have not been taken
  into account in the calculation of the two-loop Lagrangian.}.

%The result of Kong and Ravndal \cite{13} is of order
%$\frac{\alpha^2}{m^4 a^4}$:
%\begin{equation}
%\langle T^{00}\rangle\equiv u=-\frac{\pi^2}{720 a^4}-
%\frac{11}{(90)^2\cdot 30\cdot 16} \frac{\pi^4\alpha^2}{m^4 a^8} 
%.\label{3.14}
%\end{equation}
%At the two-loop level of Eq. \re{3.13}, this correction can obviously
%be neglected.
%
%  INSTEAD:
Taking the leading radiative correction to the Casimir energy into
account, Bordag, Robaschik and Wieczorek \cite{13} obtained the
result:
\begin{equation}
\langle T^{00}\rangle\equiv u=-\frac{\pi^2}{720 a^4}+\frac{1}{2560}
\frac{\alpha}{\pi} \frac{\pi^3}{ma^5}. \label{3.14}
\end{equation}
At the two-loop level of Eq. \re{3.13}, this correction can
nevertheless be neglected, since $ma\gg 1$. 

\subsection*{D. Finite Temperature.}
We begin with the one-loop correction to the effective QED Lagrangian
at finite temperature which can be decomposed according to
\begin{equation}
{\cal L}={\cal L}_{\text{M}} +{\cal L}(T=0)+\Delta{\cal L}(T),
\label{3.15}
\end{equation}
from which follows: 
\begin{equation}
Q=Q(T=0)+\Delta Q(T). \label{3.16}
\end{equation}
For purely magnetic fields, $\Delta {\cal L}$ was calculated by
Dittrich \cite{14}:
\begin{equation}
\Delta {\cal L}(B,T)=-\frac{\sqrt{\pi}}{4\pi^2}
  \int\limits_0^{\I\infty} \frac{ds}{s^{\case{5}{2}}}
  \text{e}^{-m^2 s}esB\cot esB\,\,T\left[\Theta_2(0,4\pi\text{i}sT^2)
  -\frac{1}{2T\sqrt{\pi s}} \right]. \label{3.17}
\end{equation}
Notice that in $Q=\frac{1}{2} \bigl( \partial_x^2+\partial_y^2 \bigr)
{\cal L}$, we have to differentiate with respect to $x$ and $y$. So we
must first re-introduce $x$ and $y$, differentiate and then put
$\mathbf{E}=0$, i.e., 
\begin{displaymath}
esB\cot esB\quad \longrightarrow\quad (es)^2 |y|\coth\Bigl[es\bigl(
  \sqrt{\!x^2+\!y^2}\!+x\bigr)\!^{\case{1}{2}}\Bigr]\cot\Bigl[es\bigl(
  \sqrt{\!x^2+\!y^2}\!-x\bigr)\!^{\case{1}{2}}\Bigr].
\end{displaymath}
The temperature-dependent part of the effective action charge is then
given by 
\begin{equation}
\Delta Q(B=0,T)=\frac{22}{45}\frac{\alpha^2}{m^4} \sum_{n=1}^\infty
  (-1)^n \left( \frac{m}{T} n\right)^2 K_2\left(\case{m}{T} n\right)
  .\label{3.19} 
\end{equation}
Here are some limiting cases. For low temperature we obtain:
\begin{equation}
\Delta Q(B=0,T\to 0)\simeq-\frac{22}{45}\frac{\alpha^2}{m^4}
  \sqrt{\frac{\pi}{2}} \left(\frac{m}{T} \right)^{\frac{3}{2}}
  \E^{-\frac{m}{T}} \quad\longrightarrow \quad 0^- .\label{3.20}
\end{equation}
Hence, the effective action charge is perfectly described by
\begin{displaymath}
Q(B=0,T=0)=c_1+c_2=\frac{22}{45} \frac{\alpha^2}{m^4}.
\end{displaymath}
In the high-temperature limit $\frac{T}{m}\gg 1$, the result turns out
to be
\begin{equation}
\Delta Q(T\gg m)=-\frac{22}{45} \frac{\alpha^2}{m^4} \left[
  1-\frac{k_1}{4} \frac{m^4}{T^4} +{\cal O}\left(\frac{m^6}{T^6}
  \right) \right], \qquad\quad k_1=0.123749\dots\, .\label{3.21}
\end{equation}
Therefore the complete effective action charge decreases rapidly,
$\propto \frac{1}{T^4}$:
\begin{equation}
Q(T\gg m)=Q(T=0)+\Delta Q(T\gg m) =\frac{11}{90} k_1
  \frac{\alpha^2}{T^4} +{\cal O}\left( \case{m^2}{T^6}
  \right). \label{3.22}
\end{equation}
For the LCC we need the VEV of the energy-momentum tensor which is
given by \cite{15}
\begin{equation}
\langle T_{\mu\nu} \rangle_T=\frac{\pi^2}{90} \left( N_{\text{B}}
  +\case{7}{8} N_{\text{F}} \right) T^4\,\, \text{diag} (3,1,1,1)
  .\label{3.23}
\end{equation}
For QED, we obtain:
\begin{eqnarray}
N_{\text{B}}=2\, &,&\quad N_{\text{F}}=0, \quad\text{for}\quad T\ll
  m\quad \text{(photon gas)} \nonumber\\ 
N_{\text{B}}=2\, &,&\quad N_{\text{F}}=4, \quad\text{for}\quad T\gg
  m\quad \text{(photon + } e^+e^-\text{-fermion gas)}.\nonumber 
\end{eqnarray}
So we arrive at the velocity of soft photons moving in a photon (or
photon + $e^+e^-$-) gas:
\begin{eqnarray}
\text{low T}:\qquad v&=&1-\frac{44\pi^2}{2025} \frac{\alpha^2}{m^4}\,
T^4\qquad\quad T<10^6 \, \text{K} \label{3.24}\\
\text{high T}:\qquad v&=&1-\frac{121}{8100}\, k_1 \pi^2 \alpha^2
=1-9.72\cdot 10^{-7}=\text{const.} \label{3.25}
\end{eqnarray}

\subsection*{E. Casimir Vacua at Finite Temperature.}

First we consider the low-temperature region. Here the LCC yields:
\begin{equation}
v=1+\frac{1}{(90)^2} \frac{\alpha^2}{m^4}\left(\! 11+\frac{1955}{36}
\frac{\alpha}{\pi}\right)\! \frac{\pi^2}{a^4} \left(\!
1-\frac{180\zeta (3)}{\pi^4} (Ta)^3\right) \quad \text{for}\quad
(Ta)\to 0 .\label{3.26}
\end{equation}
Neglecting $(Ta)^3$ in the low-temperature limit we end up with
Scharnhorst's result. But we do not find an additional velocity shift
$\sim T^4$, as could have been expected from Eq. \re{3.24}. This
clearly arises from the fact that none of the (quantized)
perpendicular modes can be excited at low temperature. The
$(Ta)^3$-term in Eq. \re{3.26} will become important for $(Ta)={\cal
  O}(1)$, i.e., $T>2000$ K. This shows that the Scharnhorst effect is
stable for $T<2000$ K.

For increasing temperature, we find an intermediate temperature region
characterized by the condition $1\ll Ta\ll ma$ which corresponds to
$0.2$eV $<T<0.5$MeV. This implies that $Q=Q(T=0)$ is a justified
approximation. The velocity shift in this case is given by
\begin{equation}
v=1-\frac{4\pi^2}{(45)^2} \frac{\alpha^2}{m^4}\left(\!
  11+\frac{1955}{36}\frac{\alpha}{\pi}\right)\! T^4\left(\!
  1-\frac{45\zeta (3)}{16\pi^3} \frac{1}{(Ta)^3}\right) .\label{3.27}
\end{equation}
In this limit, only the modifications caused by the blackbody
radiation become important. A term proportional to $\frac{1}{a^4}$
does not occur, since higher (perpendicular) modes have been excited.


\begin{thebibliography}{99}

\itemsep=-.2pc

\bibitem{1} S.L. Adler, Ann. Phys. (N.Y.) {\bf 67}, 599 (1971);\\
  Z. Bialynicka-Birula and I. Bialynicka-Birula, Phys. Rev. D {\bf 2},
  2341 (1970);\\
  E. Brezin and C. Itzykson, Phys. Rev. D {\bf 3}, 618 (1971).
\bibitem{2} E. Iacopini and E. Zavattini, Phys. Lett. B {\bf
    85}, 151 (1979).
\bibitem{3} R. Tarrach, Phys. Lett. B {\bf 133}, 259 (1983);\\
  G. Barton, Phys. Lett. B {\bf 237}, 559 {1990}.
\bibitem{4} K. Scharnhorst, Phys. Lett. B {\bf 236}, 354 (1990).
\bibitem{5} I.T. Drummond and S.J. Hathrell, Phys. Rev. D {\bf 22},
  343 (1980). 
\bibitem{6} J.L. Latorre, P. Pascual and R. Tarrach,
  Nucl. Phys. B {\bf 437}, 60 (1995).
\bibitem{7} Wu-yang Tsai and T. Erber, Phys. Rev. D {\bf 12}, 1132
  (1975).
\bibitem{8} G.M. Shore, Nucl. Phys. B {\bf 460}, 379 (1996).
\bibitem{9} W. Dittrich and H. Gies, Phys. Rev. D {\bf 58}, 025004
  (1998).
\bibitem{10} J. Schwinger, Phys. Rev. {\bf 82}, 664 (1951).
\bibitem{11} V.I. Ritus, JETP {\bf 42}, 774 (1976);\\
  M. Reuter, M.G. Schmidt and C. Schubert, Ann. Phys. (NY) {\bf 259},
  313 (1997).
\bibitem{12} L.S. Brown and G.J. Maclay, Phys. Rev. {\bf 184}, 1272
  (1969).  
%\bibitem{13} X. Kong and F. Ravndal, Phys. Rev. Lett. {\bf 79},545
%  (1997).
% INSTEAD:
\bibitem{13} M. Bordag, D. Robaschik and E. Wieczorek, Ann. Phys.
  (N.Y.)  {\bf 165}, 192 (1985).
\bibitem{14} W. Dittrich, Phys. Rev. D {\bf 19}, 2385 (1979).
\bibitem{15} D. Bailin and A. Love, {\em Introduction to Gauge Field
  Theory}, IOP Publishing Limited (1993).
\end{thebibliography}
\end{document}